\documentstyle[twoside,fleqn,npb,epsfig]{article}
\newcommand{\AmS}{{\protect\the\textfont2
  A\kern-.1667em\lower.5ex\hbox{M}\kern-.125emS}}

\newcommand{\gsim}{\mathrel{\vcenter
    {\hbox{$>$}\nointerlineskip\hbox{$\sim$}}}}

\title{Odderon and Photon exchange in pseudoscalar meson production}

\author{E. R. Berger \address{Theoretische Physik, Universit\"at Heidelberg,
        Philosophenweg 16, D-69120 Heidelberg, Germany}%
        \thanks{Supported by the Deutsche Forschungsgemeinschaft under 
                grant no. GRK 216/2-99}  }

\begin{document}

\begin{abstract}

We consider exclusive $\pi^0$ production in
ep-scattering. At high energies odderon and photon exchange
contribute. The photon exchange contribution 
is evaluated exactly using data for the total virtual
photon-proton absorption cross section.  
The odderon exchange contribution is calculated 
in nonperturbative QCD, using functional integral techniques 
and the model of the stochastic vacuum. For the proton we assume 
a quark-diquark structure as suggested by the
small odderon amplitude in ${pp}$ and $ { p} \bar{{p}}$
forward scattering. 
We show that odderon exchange leads to 
a much larger inelastic than elastic $\pi^0$ production cross section.
Observing our process at HERA would establish the soft
odderon. 

\end{abstract}

\maketitle

\section{Introduction}

The soft (nonperturbative) odderon $\cal{O}$ is introduced in elastic 
hadron-hadron scattering
as the $C=P=-1$ partner of the pomeron $\cal{P}$ \cite{nico1}. 
In perturbation theory the existence of $C=P=-1$ contributions
is clear (3-gluon exchange). It is even believed \cite{dolaodd1}, 
that 3-gluon exchange 
dominates the $pp$-scattering amplitude $T_{pp}$ 
for momentum transfers
$|t| \gsim 4 \; {\rm GeV}^2$. On the other hand, for 
$|t|\rightarrow 0$ it seems that odderon contributions  
play no role asymptotically or even are absent. 
An example is the measurement of 
$\rho_{p\bar{p}} = Re (T_{p\bar{p}})/Im (T_{p\bar{p}})$
at $\sqrt{s}=546$ GeV \cite{augier2}.
Together with 
$pp$ scattering data at smaller energies one deduces
$|\rho_{p p}-\rho_{p\bar{p}}| \le 0.05 $.
In other words the soft odderon
couples very weakly to the nucleon.

\begin{figure}[htb]
  \unitlength1.0cm
  \begin{center}
    \begin{picture}(15.,4.8)

      \put(-0.0,0.2){
        \epsfysize=4.7cm
        \epsffile{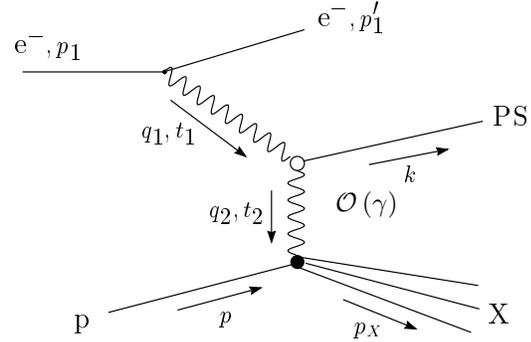}}

      \put(4.5,2.0){$\cal{O} \, (\gamma)$}

    \end{picture}
  \end{center}
  \vspace*{-1.3cm}
  \caption{Feynman diagrams for pseudoscalar meson production in $ep$
scattering at high energies with odderon (photon) exchange.}
\label{b1}
\end{figure}

Since soft elastic high energy scattering is totally dominated by the
pomeron, it is useful to look at a reaction where pomeron exchange does 
not contribute. In the following we discuss 
some results of \cite{etal} considering
exclusive $\pi^0$ production
in high energy $ep$-scattering. The $\pi^0$ is produced 
(see Fig. \ref{b1}) by
$\gamma \cal{O}$-fusion and by $\gamma \gamma$-fusion, but can not 
be produced by $\gamma \cal{P}$-fusion since the $\pi^0$ has positive 
$C$-parity. 
%
%
%
%
%
%
In Fig. \ref{b1} X stands for a proton or resonances or a sum over resonances.
Here we treat the very small $Q^2:= -q_1^2$ range. 
In the H1 experiment at HERA the kinematical cuts for this
so called photoproduction region are
\begin{eqnarray}
  && y_{{\rm min}}=0.3 \le y \le 0.7=y_{{\rm max}},
  \nonumber\\
  && 0 < Q^2 < 0.01 \, {\rm GeV^2},
  \label{cuts}
\end{eqnarray}
%
%
where, in the proton rest frame $y = (p q_1)/(p p_1)$ is the fractional
energy loss of the incoming lepton. This allowes us to use the 
equivalent photon approximation
\begin{eqnarray}
  &&\sigma_{e p} = \int_{y_{min}}^{y_{max}} \, \frac{dy}{y} \, n(y) \;
  \sigma_{\gamma {p}}
  (s_2), 
  \nonumber\\
  &&s_2=y s + (1-y) m_{{ p}}^2.
  \label{epa}
\end{eqnarray}
%
%
where $m_{p}$ 
is the nucleon mass, $\sigma_{\gamma {p}}$ is the total photoproduction
cross section for the reaction, and $n(y)$ is the equivalent photon number. 

\section{Odderon exchange}

We look at $\gamma p$-scattering in the c.m. system and
choose $\vec{q}_1$ as 3-direction. Then 
the photon and the proton have very large momenta in $\pm 3$-direction.
In the following we
discuss the cases, (i) that the proton,
which is assumed to be a quark-diquark system 
with a scalar diquark stays intact or (ii)
gets diffractively exited into the
resonances
$N(1520)$ with $J^{P}={\frac{3}{2}}^- $ 
and $N(1535)$ with $J^{P}={\frac{1}{2}}^- $,
described as exited quark-diquark systems. 

\begin{figure}[ht]
 
\vspace*{0.0cm}
 
\hspace{1cm}
\epsfysize=3.2cm
\centerline{\epsffile{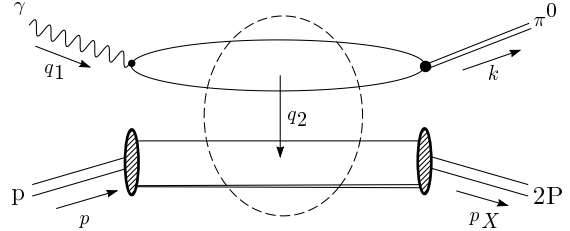}}
 
\vspace*{-0.5cm}
 
\caption{$\pi^0$ production in $\gamma p$-scattering. 
The dashed circle indicates the nonperturbative interaction
(odderon exchange)
of the colour dipoles.} 

\label{gampbild}
\end{figure}

When considering unpolarised cross sections,
summed over both resonances in case (ii), 
the quark spin degree of freedom becomes
irrelevant and the calculation reduces to one where 
a spinless state stays intact  or is
exited to a 2P resonance. 
For (ii) the helicity amplitudes are \cite{etal}
\begin{eqnarray}
 &&T(s_2,t_2)_{\lambda,\lambda_{\gamma}} = 
 - 2is_2 \int \, d^2b\,e^{i\vec{q_2}_T \vec{b}}\,
 \hat{J}_{\lambda,\lambda_{\gamma}} (\vec b),
 \nonumber\\
 && \hat{J}(\vec{b})_{\lambda,\lambda_{\gamma}} = 
 \int \frac{d^2 r_1}{4\pi} dz
 \int \frac{d^2 r_2}{4\pi} 
\nonumber\\
 &&\hphantom{\hat{J}(\vec{b})}
\times
\sum_{f,h_1, h_2} 
 \Psi^{*\, \pi^0}_{f h_1 h_2}(\vec{r}_1,z)
 \Psi^{\gamma}_{\lambda_{\gamma},\,f h_1 h_2} (\vec{r}_1,z)
 \nonumber\\
 &&\hphantom{\hat{J}(\vec{b})}
 \times
 \Psi^{*\, {\rm 2P}}_{\lambda} (\vec{r}_2)
 \Psi^{p}(\vec{r}_2) \; \;
 \tilde{J}(\vec b, \vec{r}_1,z, \vec{r}_2).
\label{psamp}
\end{eqnarray}
Here $\lambda_{\gamma} (\lambda)$ is the helicity of the photon (2P state)
and
$z$ is the momentum fraction of the photon, carried by the quark.
The physical picture arising from (\ref{psamp}) is the following 
(Fig. \ref{gampbild}).
The photon fluctuates into a $q\bar{q}$ pair, described by $\Psi^{\gamma}$.
By soft colour interaction (odderon exchange), calculated 
from the functional integral of two lightlike Wegner-Wilson loops 
($\tilde{J}$) the $q \bar{q}$ pair turns into a $\pi^0$ and the proton
either stays intact or gets
exited (described by $\Psi^{\rm 2P}$).
%
%
%
%
%

Now, 
when the proton stays intact there occurs in (\ref{psamp})
the quark-diquark density
($\Psi_{\lambda}^{\rm 2P} \rightarrow \Psi^p$). 
But, since this 
density is  symmetric under a parity transformation 
whereas the odderon coupling changes
sign, there is a cancelation when we integrate 
over all angles. On the other hand when the
proton gets exited to a negative parity state like the resonances
$N(1520)$ and $N(1535)$ there is no cancellation (the overlap is odd
under parity) and the
odderon couples to the nucleon without any restriction 
\cite{etal,donaru}. Thus in our model the odderon couples only if  
breakup of the proton occurs.

\subsection{Results}

\begin{figure}[htb]
  \unitlength1.0cm
  \begin{center}
    \begin{picture}(15.,4.8)
      \put(1.0,0.0){
        \epsfysize=6.5cm

        \epsffile{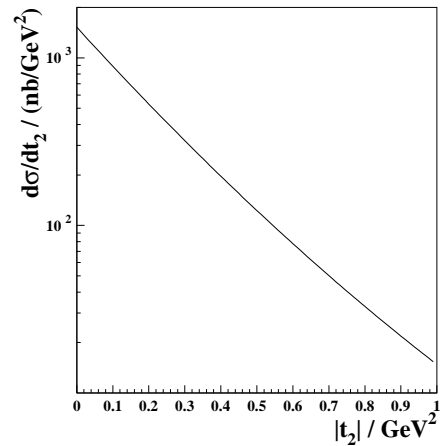}}
    \end{picture}
  \end{center}
  \vspace*{-1.3cm}
  \caption{The differential cross section 
     $d\sigma_{\gamma { p}}^{\cal{O}}  / dt_2$ 
  of the process
  $\gamma { p} \to \pi^0 \{{\rm 2P}\}$ as a function of $t_2$.}
  \label{msvbild}
\end{figure}

\begin{figure}[htb]
  \unitlength1.0cm
  \begin{center}
    \begin{picture}(15.,4.4)
      \put(0.5,-0.2){
        \epsfysize=6.5cm

        \epsffile{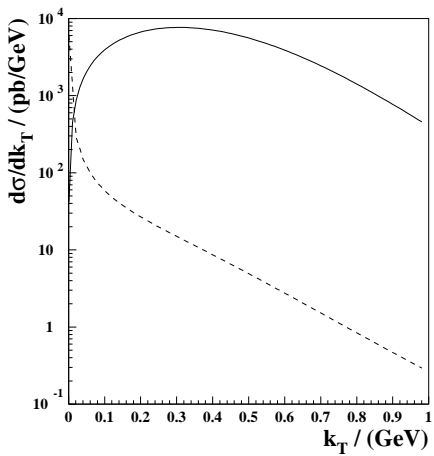}}
    \end{picture}
  \end{center}
  \vspace*{-1.1cm}

  \caption{The $k_T$ distribution in pion production
from the 2P resonance channels for 
odderon exchange (solid line)
compared to the
complete electromagnetic result (dashed line). Interference 
contributions are not taken into account.}
\label{k_Tmsv}
\end{figure}

In Fig. \ref{msvbild} we show our result for 
the differential cross section in $\gamma p$-scattering,
$d \sigma_{\gamma p} / d t$. 
%
%
%
%
%
%
The slope of $d \sigma_{\gamma p}^{\cal{O}} / d t$ 
at $t=0$ is around $5/{\rm GeV}^2$. 
The integrated cross section is \cite{etal}:
\begin{eqnarray}
\sigma_{\gamma \, {\rm p}}^{\cal{O}}
(\gamma {\rm p} \to \pi^0 \{2{\rm P}\}) = 294 \;{\rm nb}.
\label{oddcross}
\end{eqnarray}
As this photoproduction cross section is constant, i.e. independent of $s_2$,
the EPA conversion to electroproduction can be achieved by simply multiplying
it with a constant $c_{{\rm EPA}} = 0.0136$ corresponding to the $y$ 
integration 
in (\ref{epa}). This gives
\begin{eqnarray}
\sigma^{\cal{O}}
({ e p} \to { e} \pi^0 \{{\rm 2P}\}) =  4010 \;{\rm pb}.
\end{eqnarray}
In the following we consider only $ep$-scattering.
An experimentally prefered observable is the $k_T$ spectrum,
the transverse momentum distribution of the $\pi^0$ with respect 
to the beam direction. 
At HERA $k_T$ distributions can be measured for values of $k_T$ 
greater than $O$(0.1 GeV). 
The photoproduction cuts of (\ref{cuts}) restrict the 
transverse 
momentum of the incident photon to be smaller than $O$(0.1 GeV), so in 
our case there is 
practically no distinction between the beam axis and the photon
axis. The $k_T$ spectrum is displayed in Fig. \ref{k_Tmsv} 
%
%
%
%
%
%
\section{Photon exchange}
In this section we consider PS production mediated by photon
rather than odderon exchange (Fig. \ref{b1}).
Again the proton, now hit by the photon, is allowed to go 
into some hadron final state X. The process has been 
calculated in \cite{kina} for X being a proton. In \cite{etal}
we have calculated the cross section, when X is a hadron final state
in the invariant mass ($M_X$) range $1.11 \le M_X \le 1.99$ GeV.
In Fig. \ref{k_Tmsv} we show  the 
distribution of the pion's $k_T$ summed over the elastic and 
all inelastic channels for the electromagnetic exchange. As we can see there 
the photon exchange is larger than the odderon exchange only
for very small $k_T$. For $k_T \gsim 0.1$ GeV the Odderon exchange
dominates by orders of magnitude. 
The integrated cross section is \cite{etal}:
\begin{eqnarray}
\sigma^{\gamma}
({ e p} \to { e} \pi^0 \{{\rm p+X}\}) =  80.1 \;{\rm pb}.
\end{eqnarray}
which has to be compared with the odderon cross 
section (\ref{oddcross}).

In addition we have estimated in \cite{etal} contributions 
from reggeon ($\omega$) exchange and  from a 
possible background process, namely an additional
final state photon emitted from the $\pi^0$ vertex (Fig. \ref{b1}),
where we have used realistc cuts. Both contributions are of the same
order of magnitude as the 
photon exchange contribution. 

\vskip 0.4cm

\noindent
The conclusion is: With the large odderon cross section
as given in our model
the process should be observable at HERA. 
This would establish the soft odderon as an exchange-object
in high energy scattering on an equal footing with the soft pomeron.


\end{document}